\documentclass[10pt]{article}
\usepackage{amsmath}
\usepackage{graphicx}
\usepackage{color}
\usepackage{orcidlink}
\usepackage{hyperref}
\hypersetup{colorlinks=true, linkcolor=blue, citecolor=blue, urlcolor=blue}
\begin{document}
\baselineskip=17 pt

\begin{center}
    {\large{\bf Topologically Charged Rotating Wormhole}}
\end{center}

\vspace{0.5cm}

\begin{center}
     {\bf Faizuddin Ahmed\orcidlink{0000-0003-2196-9622}}\footnote{\bf faizuddinahmed15@gmail.com}\\
    \vspace{0.1cm}
    {\it Department of Physics, University of Science \& Technology Meghalaya,Ri-Bhoi, 793101, India}
\end{center}

\vspace{0.5cm}

\begin{abstract}
In this article, we present a stationary metric ansatz to describe a rotating traversable wormhole in the presence of the topological defect produced by a global monopole charge. This particular rotating space-time is referred to as the topologically charged rotating Schwarzschild-Klinkahmer wormhole. Our study involves the analysis of geodesic motion for test particles and photon rays in the context of this topologically charged rotating traversable wormhole. We aim to analyze the effects of global monopole charge and other parameters on the outcomes of this investigation. Additionally, we explore the matter-energy distribution within this rotating wormhole, considering it as a non-vacuum solution of Einstein's field equation. Notably, we demonstrate that the energy density of the matter content satisfies the criteria of the weak energy condition.
\end{abstract}

{\bf keywords}: Modified theories of gravity; wormhole; global monopoles; energy condition

\section{Introduction}

Wormholes, intriguing hypothetical constructs within the fabric of space-time, possess non-trivial topologies that have the potential to link distant regions of a single universe or even bridge between separate universes. These enigmatic passages consist of two distinct entry points known as ``mouths", connected by a slender corridor referred to as the ``throat". The fundamental configuration of a wormhole comprises these dual mouths joined by the narrow throat. It's important to note that wormholes don't emerge as direct predictions of general relativity or other established theories of gravity. Rather, they manifest as intricate space-time formations that could potentially emerge within the curvature of space-time. This possibility encompasses a broad spectrum of gravitational models. The mechanisms governing their formation and stability vary depending on the specific theory of gravity being considered. These mechanisms often pose challenging issues and complexities, thereby complicating the assessment of the existence and viability of wormholes in the universe. Despite the hurdles, one cannot definitively dismiss the prospect of wormholes within our cosmic landscape. Numerous solutions describing wormhole space-times have been formulated, encompassing scenarios both with and without a cosmological constant, as outlined in reference \cite{MV}. Nevertheless, a significant apprehension accompanying these space-time constructs is their status as exact solutions to field equations, frequently entailing breaches of one or more energy conditions. The violation of foundational principles like the weak energy condition (WEC) and the null energy condition (NEC) represents a central concern within the realm of wormhole space-times.

Despite the challenges posed by these energy condition violations, the exploration of wormholes has ignited substantial scientific interest, catalyzing ongoing research across diverse interconnected fields. The theoretical pursuit of understanding these extraordinary space-time geometries continues to broaden our comprehension of not only general relativity but also other fundamental facets of physics. Through these investigations, our insights into the intricate nature of space-time itself, as well as the boundaries and possibilities of theoretical physics, are continually enriched. The field of non-vacuum wormhole space-times has made remarkable strides, surmounting the longstanding hurdle of exotic matter. Notably, the acknowledgement that vacuum solutions of the field equations inherently align with energy conditions has been instrumental. Building upon these foundational insights, a pivotal contribution was made in Ref. \cite{FRK}, where a traversable wormhole space-time, either non-vacuum or vacuum, was introduced. Importantly, this innovative construction demonstrated adherence to both the weak and null energy conditions, marking a significant advancement. In Ref. \cite{FRK3}, a multiple vacuum wormhole solution was proposed by Klinkhamer. 

The non-vacuum space-time describing the Klinkhamer wormhole is described by the following line-element \cite{FRK}
\begin{equation}
    ds^2=-dt^2+\Big(1+\frac{b^2}{\xi^2}\Big)^{-1}\,d\xi^2+(\xi^2+\lambda^2)\,(d\theta^2+\sin^2 \theta\,d\phi^2),
    \label{K1}
\end{equation}
where $-\infty < (t, \xi) < +\infty$ and other coordinates are in the usual ranges, and $\lambda, b$ are arbitrary constants. For $\lambda^2=b^2$, this metric (\ref{K1}) becomes a vacuum defect wormhole space-time \cite{FRK3}. However, in Ref. \cite{FA}, we presented non-rotating topologically charged wormholes metric (like Klinkhamer, Schwarzschild-Klinkhamer and it's generalization), which are non-vacuum solutions of the Einstein's field equations obeying the weak and null energy conditions. In ref. \cite{FA2}, we presented a topologically charged four-dimensional wormhole called topologically charged Schwarzschild-Simpson-Visser wormhole satisfying the energy condition The topologically charged Schwarzschild-Klinkhamer-wormhole metric is described by the following line-element \cite{FA}
\begin{equation}
    ds^2=-\Big(1-\frac{2\,M}{\sqrt{\xi^2+b^2}}\Big)\,dt^2+\Big(1+\frac{b^2}{\xi^2}\Big)^{-1}\,\Big(1-\frac{2\,M}{\sqrt{\xi^2+b^2}}\Big)^{-1}\,\frac{d\xi^2}{\alpha^2}+(\xi^2+b^2)\,(d\theta^2+\sin^2 \theta\,d\phi^2),
    \label{K2}
\end{equation}
where $b>2\,M$ is arbitrary constant.

Using the above space-time, we discussed geodesics motions of the test particles and bending of photon ray in details. One can see that in the limit $M \to 0$, this metric (\ref{K2}) reduces to a topologically charged non-vacuum wormhole space-time \cite{FA}. And in the particular case $b \to 0$ and $\alpha \to 1$, it recovers the Schwarzschild black hole solution. Notably, it's important to highlight that in a separate study in \cite{JCF}, an analysis is presented indicating that smooth metrics can effectively conceal the presence of thin shells. Furthermore, an additional commentary on defect wormholes is provided in \cite{BGV}. The defect wormhole solution remains a pertinent topic within the realm of nonstandard general relativity. In this context, the presence of a defect at the point $x=0$ signifies an inherent imperfection within the fabric of space-time. It is at this juncture that the conventional elementary-flatness condition, which characterizes smooth and unblemished space-time, fails to hold true. Regrettably, it seems that this crucial aspect has been inadvertently overlooked in the more recent papers, namely, the works of \cite{JCF} and \cite{BGV}.

Rotating wormholes of stationary and axially symmetric model has been investigated only in a few works. The first such investigation was done by Teo \cite{ET}, an extension of Morris-Thorne traversable wormhole. The canonical metric form of Teo wormhole is described by the following line-element \cite{ET}
\begin{eqnarray}
    ds^2=-N^2\,dt^2+\Big(1-\frac{b}{r}\Big)^{-1}\,dr^2+r^2\,K^2\,\Big[d\theta^2+\sin^2 \theta\,\big(d\varphi-\omega\,dt\big)^2\Big],
    \label{K3}
\end{eqnarray}
where the author chosen $N, b, K$ and $\omega$ are functions of $r$ and $\theta$ rather than simply parameters such that this metric is regular on the symmetry axis $\theta=0, \pi$.

In Ref. \cite{VMK}, a slowly rotating spherically symmetric wormhole was constructed, which is a vacuum solution of the field equations. The space-time is described by the following line-element
\begin{eqnarray}
    ds^2=e^{2\,\Phi (\rho)}\,dt^2-d\rho^2-r^2\,\Big[d\theta^2+\sin^2 \theta \,\big(d\varphi^2+2\,h\,d\varphi\,dt\big)\Big],
    \label{K4}
\end{eqnarray}
where $h=h(\rho, \theta)$ has the sense of angular velocity of the local inertial frame and $r=r(\rho)$. A few other investigations of rotating wormhole space-times are in Refs. \cite{ss1,ss2,ss3,ss4,ss5,hh1,hh2,hh3,hh4,hh5}.

Topological defects are theoretical entities believed to have formed during the early stages of the universe's evolution, as extensively discussed in the review articles \cite{kk1,kk2,kk3,AV,kk51,kk52,kk5,kk6}. These defects could have emerged during one of the early phase transitions in particle physics models, such as those associated with the breaking of grand unification symmetry or spontaneously broken gauge theories, as proposed in numerous field theory models \cite{kk7,kk8,kk9}. Among these defects, global monopoles take the form of spherically symmetric objects arising from self-coupling triplet of scalar fields $\phi^{a}$. These scalar fields undergo a spontaneous breaking of the global $O(3)$ gauge symmetry, resulting in structures akin to cosmic strings with $U(1)$ symmetry. However, global monopoles exhibit distinct properties than the cosmic strings. Notably, studies have demonstrated that these topological defects possess a negative gravitational potential \cite{kk12}. The concept of global monopole charge has also been found relevance in cosmological contexts, as evidenced by the research in \cite{kk13,kk14,kk15}. Moreover, recent investigations have explored the existence of wormholes within the Milky Way galaxy, taking into account the presence of global monopole charge, as discussed in \cite{kk16}.

In this article, our objective is to introduce a metric ansatz that describes a topologically charged rotating Schwarzschild-Klinkhamer wormhole space-time. We will explore the geodesic motions of test particles and photon rays in various directions around this rotating wormhole and analyze the outcomes. Furthermore, we discuss various physical quantities associated with the curvature tensor, such as the Kretschmann scalar, the quadratic Ricci invariant, and the Ricci scalar. We demonstrate that these quantities are finite throughout the space-time and are influenced by the global monopole charge. Finally, we study the matter-energy content of this rotating wormhole and show that the energy density satisfies the weak energy condition, thus indicating an example of rotating wormhole without the need for exotic matter.

The paper is structured as follows: In Section 2, we conduct an analysis of a topologically charged rotating traversable wormhole. Within this section, we examination the geodesic motions of test particles around this wormhole, with a detailed discussion of the outcomes provided in Sub-section 2.1. Additionally, we explore the matter-energy content associated with this rotating wormhole, recognizing it as a non-vacuum solution of the Einstein's field equations, and present our findings in Sub-section 2.2. Moving on to Section 3, we present our results and discuss the outcomes and insights derived from our examination of this rotating wormhole. Throughout the paper, we choose the systems of units, where $c=1=8\,\pi\,G$.

\section{Analysis of a topologically charged rotating wormhole}

In this section, we consider the following ansatz of a stationary topologically charged rotating wormhole of the metric (\ref{K2}) by the following line-element 
\begin{eqnarray}
ds^2&=&-A\,dt^2+B\,\Big(1+\frac{b^2}{\xi^2}\Big)\,\frac{d\xi^2}{\alpha^2}+\Lambda\,d\theta^2-2\,H\,dt\,d\phi+K\,d\phi^2\nonumber\\
&=&-\Bigg(1-\frac{2\,M\,\sqrt{\xi^2+b^2}}{\xi^2+b^2+a^2\,\cos^2 \theta}\Bigg)\,dt^2+\Bigg(\frac{\xi^2+b^2+a^2\,\cos^2 \theta}{\xi^2+b^2+a^2-2\,M\,\sqrt{\xi^2+b^2}}\Bigg)\times\nonumber\\
&&\Big(1+\frac{b^2}{\xi^2}\Big)^{-1}\,\frac{d\xi^2}{\alpha^2}+(\xi^2+b^2+a^2\,\cos^2 \theta)\,d\theta^2-\Bigg(\frac{4\,M\,a\,\sqrt{\xi^2+b^2}\,\sin^2 \theta}{\xi^2+b^2+a^2\,\cos^2 \theta}\Bigg)\,dt\,d\phi\nonumber\\
&+&\Bigg(\xi^2+b^2+a^2+\frac{2\,M\,a^2\,\sqrt{\xi^2+b^2}\,\sin^2 \theta}{\xi^2+b^2+a^2\,\cos^2 \theta}\Bigg)\,\sin^2 \theta\,d\phi^2,
\label{1}
\end{eqnarray}
with $M \neq 0$ represents mass of the objects, $a=J/M$ represents spin angular momentum per unit mass, and the parameter $b$ is strictly a non-zero positive parameter, $b>2\,M$. In the particular case $a=0$, this metric (\ref{1}) reduces to a topologically charged non-rotating SK-wormhole space-time (\ref{K2}) which was discussed in details in Ref. \cite{FA}. The coordinates are in the ranges 
\begin{equation}
-\infty < t < +\infty,\quad -\infty < \xi < +\infty,\quad 0 < \theta < \pi,\quad 0 \leq \phi < 2\,\pi. \label{coordinate}    
\end{equation}

In the particular case $b=0$ (which is restricted here), the metric (\ref{1}) reduces to the following form
\begin{eqnarray}
ds^2&=&-\Bigg(1-\frac{2\,M\,\xi}{\xi^2+a^2\,\cos^2 \theta}\Bigg)\,dt^2+\Bigg(\frac{\xi^2+a^2\,\cos^2 \theta}{\xi^2+a^2-2\,M\,\xi}\Bigg)\,\frac{d\xi^2}{\alpha^2}+(\xi^2+a^2\,\cos^2 \theta)\,d\theta^2\nonumber\\
&&-\Bigg(\frac{4\,M\,a\,\xi\,\sin^2 \theta}{\xi^2+a^2\,\cos^2 \theta}\Bigg)\,dt\,d\phi
+\Bigg(\xi^2+a^2+\frac{2\,M\,a^2\,\xi\,\sin^2 \theta}{\xi^2+a^2\,\cos^2 \theta}\Bigg)\,\sin^2 \theta\,d\phi^2
\label{2}
\end{eqnarray}
which is a non-vacuum solution of the field and becomes vacuum one, the Kerr-like rotating metric \cite{RPK} in the limit $\alpha \to 1$. Our aim is to analyze the above considered topologically charged wormhole under the case $b \neq 0$.

In the limit $M \to 0$, the metric (\ref{1}) reduces to the following form
\begin{eqnarray}
      ds^2&=&-dt^2+\Big(\frac{\xi^2+b^2+a^2\,\cos^2 \theta}{\xi^2+b^2+a^2}\Big)\,\Big[\Big(1+\frac{b^2}{\xi^2}\Big)^{-1}\,\frac{d\xi^2}{\alpha^2}+(\xi^2+b^2+a^2)\,d\theta^2\Big]\nonumber\\
      &&+(\xi^2+b^2+a^2)\,\sin^2 \theta\,d\phi^2.
    \label{a1}
\end{eqnarray}
One can see that for zero rotation, $a \to 0$, the metric (\ref{a1}) reduces to a topologically charged Klinkhamer wormhole non-vacuum space-time \cite{FA} and subsequently the Klinkhamer vacuum defect wormhole \cite{FRK3} for $\alpha \to 1$.

Now, we recall the metric component $g_{\phi\phi}$ for the space-time (\ref{1}) given by
\begin{equation}
    g_{\phi\phi}=\Bigg[\xi^2+b^2+a^2+\frac{2\,M\,a^2\,\sqrt{\xi^2+b^2}\,\sin^2 \theta}{\xi^2+b^2+a^2\,\cos^2 \theta}\Bigg]\,\sin^2 \theta
    \label{a2}
\end{equation}
which is always positive except on the axis $\theta=0, \pi$, that is, $g_{\phi\phi}>0$ indicating that the space-time (\ref{1}) satisfies the causality condition, and thus, no closed time-like curves will be formed unlike the Kerr metric \cite{RPK}. Note that in the Kerr metric, CTCs will disappear in the limit $M \to 0$. 

At $\xi=0$ in the equatorial plane $\theta=\pi/2$, from (\ref{a2}), we obtain 
\begin{equation}
    g_{\phi\phi}|_{\xi=0,\theta=\pi/2}=\Big(b^2+a^2+\frac{2\,M\,a^2}{b}\Big)>0
    \label{aa2}
\end{equation}
which is positive since $b>0$, and thus, no closed time-like curves will appear.

The space-time (\ref{1}) can be expressed as 
\begin{eqnarray}
    &&ds^2=-dt^2+\Bigg[\frac{\xi^2+b^2+a^2\,\cos^2 \theta}{\xi^2+b^2+a^2-2\,M\,\sqrt{\xi^2+b^2}}\Bigg]\,\Big(1+\frac{b^2}{\xi^2}\Big)^{-1}\,\frac{d\xi^2}{\alpha^2}+(\xi^2+b^2+a^2\,\cos^2 \theta)\,d\theta^2\nonumber\\
    &&+(\xi^2+b^2+a^2)\,\sin^2 \theta\,d\phi^2+\frac{2\,M\,\sqrt{\xi^2+b^2}}{\xi^2+b^2+a^2\,\cos^2 \theta}\,(dt-a\,\sin^2 \theta\,d\phi)^2.
    \label{a3}
\end{eqnarray}

On the axis of rotation $(\theta=0, \theta=\pi)$, this line-element (\ref{a3}) reduces to
\begin{eqnarray}
     ds^2=-\Bigg[1-\frac{2\,M}{\sqrt{\xi^2+b^2}\Big(1+\frac{a^2}{\xi^2+b^2}\Big)}\Bigg]dt^2+\Bigg[1-\frac{2\,M}{\sqrt{\xi^2+b^2}\Big(1+\frac{a^2}{\xi^2+b^2}\Big)}\Bigg]^{-1}\Big(1+\frac{b^2}{\xi^2}\Big)^{-1}\frac{d\xi^2}{\alpha^2}.
    \label{aa3}
\end{eqnarray}

On the equatorial plane $(\theta=\pi/2)$,  the line-element (\ref{a3}) reduces to
\begin{eqnarray}
    ds^2=-dt^2+\frac{\Big(1+\frac{a^2}{\xi^2+b^2}-\frac{2M}{\sqrt{\xi^2+b^2}}\Big)^{-1}}{\Big(1+\frac{b^2}{\xi^2}\Big)}\,\frac{d\xi^2}{\alpha^2}+(\xi^2+b^2+a^2)d\phi^2+\frac{2M}{\sqrt{\xi^2+b^2}}(dt-a\,d\phi)^2.
    \label{aaa3}
\end{eqnarray}

Transforming to a new coordinate via $r=\sqrt{\xi^2+b^2}$ into the space-time (\ref{a3}) results 
\begin{eqnarray}
    &&ds^2=-dt^2+\Big(\frac{r^2+a^2\,\cos^2 \theta}{r^2+a^2-2\,M\,r}\Big)\,\frac{dr^2}{\alpha^2}+(r^2+a^2\,\cos^2 \theta)\,d\theta^2+(r^2+a^2)\,\sin^2 \theta\,d\phi^2\nonumber\\
    &&+\frac{2\,M\,r}{r^2+a^2\,\cos^2 \theta}\,(dt-a\,\sin^2 \theta\,d\phi)^2
    \label{a4}
\end{eqnarray}
which looks for $\alpha \to 1$ though similar to the Kerr-like metric in Boyer and Lindquist coordinates \cite{RPK} but the coordinate $r$ here is in the range $r \in [b>2\,M, \infty)$ unlike the Kerr one. To check it, the metric components $g_{tt}$ and $g_{rr}$ for the space-time (\ref{a4}) are 
\begin{eqnarray}
    g_{tt}|_{\xi=0,\theta=\pi/2}=-\Big(1-\frac{2\,M}{b}\Big), \quad g_{rr}|_{\xi=0,\theta=\pi/2}=\frac{1}{\alpha^2}\,\frac{b^2}{b\,(b-2\,M)+a^2}.
    \label{a5}
\end{eqnarray}
From the above discussion, we see that the metric component $g_{tt}<0$ and $g_{rr}>0$ provided the parameter $b>2\,M$ to prevent the formation of event horizons so that it represents a wormhole space-time. Noted that our metric (\ref{1}) or (\ref{a4}) is a non-vacuum solution of the Einstein's field equations, $G_{\mu\nu}=T_{\mu\nu}$ which will be discussed here in later on. 

Below, we investigate the geodesic motions of particles (either time-like or photon rays) around this topologically charge wormhole in various direction and analyze the effects of global monopole charge ($\alpha$), wormhole parameter ($b$), rotation parameter ($a$) including mass $M$ of the objects.

\subsection{\bf Motion of photon rays in different directions}

In this part, we discuss motions of photon ray in the equatorial plane $\theta=\pi/2$ in various direction as well as geodesics of test particles around this topologically charged rotating SK-wormhole (\ref{1}).

\vspace{0.2cm}
{\bf Case 1: Photon ray moving in the $\phi$-direction:}
\vspace{0.2cm}

We consider the motion of photon ray in the $\phi$-direction. In that case $d\xi=0$, therefore, for the space-time (\ref{1}) we have
\begin{equation}
    -A-2\,H\,\frac{d\phi}{dt}+K\,\Big(\frac{d\phi}{dt}\Big)^2=0
    \label{13}
\end{equation}
which is a quadratic equation whose solution is given by
\begin{equation}
    \Big(\frac{d\phi}{dt}\Big)_{\pm}=\frac{H}{K}\pm \sqrt{\frac{H^2}{K^2}+\frac{A}{K}},
    \label{14}
\end{equation}
where
\begin{eqnarray}
    \frac{H}{K}=\frac{2 M a}{\sqrt{\xi^2+b^2}\,\Big(\xi^2+b^2+a^2+\frac{2 M a^2}{\sqrt{\xi^2+b^2}}\Big)},\quad
    \frac{A}{K}=\frac{\Big(1-\frac{2 M}{\sqrt{\xi^2+b^2}}\Big)}{\Big(\xi^2+b^2+a^2+\frac{2 M a^2}{\sqrt{\xi^2+b^2}}\Big)}.
    \label{15}
\end{eqnarray}
At the wormhole throat $\xi=0$, we obtain
\begin{eqnarray}
    \frac{d\phi}{dt}|_{\xi=0, a\neq 0}=\Big(b^2+a^2+\frac{2\,M\,a^2}{b}\Big)^{-1}\,\Bigg[\frac{2\,M\,a}{b}\pm\,\sqrt{b^2+a^2-2\,M\,b}\Bigg].
    \label{16}
\end{eqnarray}
For zero rotation, that is, $a=0$, it becomes $\frac{d\phi}{dt}|_{\xi=0}=\pm\,\frac{1}{b}\sqrt{1-\frac{2\,M}{b}}$, where $b>2\,M$.

In the limit $M \to 0$ and non-rotating case, we obtain from (\ref{14})
\begin{eqnarray}
    \frac{d\phi}{dt}|_{M \to 0, a \neq 0}=\pm\,\frac{1}{\sqrt{\xi^2+b^2+a^2}}.
    \label{16a}
\end{eqnarray}

Thus, we see that when the photon ray propagating along the $\phi$-direction, the quantity $\frac{d\phi}{dt}$ is influenced by the wormhole parameter $b$, the rotation parameter $a$, and the mass $M$ of the objects.

\vspace{0.2cm}
{\bf Case 2: Photon ray moving in the tangential direction:}
\vspace{0.2cm}

The physical velocities of light moving in the tangential direction is defined by
\begin{equation}
    v_{\pm}=\sqrt{|g_{\phi\phi}|}\,\Big(\frac{d\phi}{dt}\Big)_{\pm}=\Bigg(\xi^2+b^2+a^2+\frac{2\,M\,a^2}{\sqrt{\xi^2+b^2}}\Bigg)^{1/2}\,\Big(\frac{d\phi}{dt}\Big)_{\pm},
    \label{17}
\end{equation}
where $\Big(\frac{d\phi}{dt}\Big)_{\pm}$ is given by equation (\ref{14}).

At the wormhole throat, $\xi=0$, we obtain 
\begin{equation}
    v|_{\xi=0, a\neq 0}=\Bigg(b^2+a^2+\frac{2\,M\,a^2}{b^2}\Bigg)^{-1/2}\,\Bigg[\frac{2\,M\,a}{b}\pm\,\sqrt{b^2+a^2-2\,M\,b}\Bigg].
    \label{17a}
\end{equation}
For zero rotation, $a=0$, it becomes $v|_{\xi=0}=\pm\,\sqrt{1-\frac{2\,M}{b}}$, where $b>2\,M$.

Also, in the limit $M \to 0$, we obtain from (\ref{17})
\begin{equation}
    v|_{M \to 0, a\neq 0}=\pm\,1.
    \label{17b}
\end{equation}

Thus, the velocities of light rays moving along the tangential direction is influenced by rotation parameter $a$, the wormhole parameter $b$ including mass M of the objects.

\vspace{0.2cm}
{\bf Case 3: Photon ray moving in the radial direction:}
\vspace{0.2cm}

For the photon ray moving in the radial direction, $d\phi=0$. Thus, we obtain from (\ref{1}) after transforming $r=\sqrt{\xi^2+b^2}$, where $r \in [b>0, \infty)$ as
\begin{equation}
    B\,\frac{dr^2}{\alpha^2}=A\,dt^2\Rightarrow \frac{dr}{dt}=\pm\,\alpha\,\sqrt{\Big(1-\frac{2\,M}{r}\Big)\Bigg(1+\frac{a^2}{r^2}-\frac{2\,M}{r}\Bigg)}.
    \label{18}
\end{equation}

In the limit $M \to 0$, from (\ref{18}) we obtain 
\begin{equation}
    \frac{dr}{dt}|_{M\to 0, a\neq 0}=\pm\,\frac{\alpha}{r}\,\sqrt{r^2+a^2}.
\end{equation}

At the wormhole throat, $\xi=0$, from (\ref{18}) we obtain 
\begin{equation}
\frac{dr}{dt}|_{\xi=0, a \neq 0}=\pm\,\alpha\,\sqrt{\Big(1-\frac{2\,M}{b}\Big)\Bigg(1+\frac{a^2}{b^2}-\frac{2\,M}{b}\Bigg)},
\label{19}
\end{equation}
where $b>2\,M$.

Thus one can see that when photon rays moving along the radial direction, the quantity $\frac{dr}{dt}$ is influenced by the topological defect of global monopole charge ($\alpha$). Furthermore, different parameters ($a, b$) including mass $M$ of the objects also influenced on it.

Lastly, we study the geodesics motions of test particles around this topologically charged rotating wormhole in the equatorial plane defined by $\theta=\pi/2$. The Lagrangian of the system in the equatorial plane $\theta=\frac{\pi}{2}$ for the space-time (\ref{1}) is given by \cite{FA}
\begin{equation}
    \mathcal{L}=\frac{1}{2}\,\Big[-A\,\dot{t}^2+B\,\Big(1+\frac{b^2}{\xi^2}\Big)^{-1}\,\frac{\dot{\xi}^2}{\alpha^2}-2\,H\,\dot{t}\,\dot{\phi}+K\,\dot{\phi}^2\Big].
    \label{4}
\end{equation}
From above, we have two constant of motions defined by
\begin{eqnarray}
    E=A\,\dot{t}+H\,\dot{\phi},\quad L=-H\,\dot{t}+K\,\dot{\phi}
    \label{5}
\end{eqnarray}
Simplification of the above equation results
\begin{equation}
    \dot{\phi}=E\,\Big(\frac{H+\beta\,A}{H^2+A\,K}\Big)\quad,\quad \dot{t}=E\,\Big(\frac{K-\beta\,H}{H^2+A\,K}\Big),
    \label{6}
\end{equation}
where $\beta=L/E$ is the impact parameter.

For geodesics motions (either time-like or null) from Eq. (\ref{4}) after transforming $r=\sqrt{\xi^2+b^2}$, we obtain
\begin{eqnarray}
    \dot{r}^2=\frac{\alpha^2}{B}\Big[\varepsilon+\frac{E^2}{(H^2+A K)^2}\Big\{A (K-\beta H)^2+2 H (H+\beta A)(K-\beta H)-K(H+\beta A)^2\Big\}\Big],
    \label{7}
\end{eqnarray}
where $\varepsilon=0$ for null geodesics and $-1$ for time-like geodesics, and $r \in [b>2\,M, \infty)$ with different functions
\begin{eqnarray}
    A=\Big(1-\frac{2\,M}{r}\Big),\quad B=\frac{r^2}{r^2+a^2-2\,M\,r},\quad H=\frac{2\,M\,a}{r}\quad K=r^2+a^2+\frac{2\,M\,a^2}{r}.
    \label{8}
\end{eqnarray}
Writing Eq. (\ref{7}) as $\dot{r}^2=V_{eff} (r)$, where we introduce an effective potential $V_{eff}$ given by
\begin{equation}
    V_{eff}=\frac{\alpha^2}{B}\,\Big[\varepsilon+\frac{E^2}{(H^2+A\,K)^2}\,\Big\{A\,(K-\beta\,H)^2+2\,H\,(H+\beta\,A)(K-\beta\,H)-K\,(H+\beta\,A)^2\Big\}\Big].
    \label{9}
\end{equation}

For zero angular momentum observer (ZAMO), $L=0$ leads to the following quantity
\begin{equation}
    \frac{\dot{\phi}}{\dot{t}}=\omega=\frac{H}{K}=\frac{2\,M\,a}{r^3+a^2\,r+2\,M\,a^2}
    \label{10}
\end{equation}
we called angular velocity (also known as coordinate rate of rotation) of the inertial frame relative to the reference frame. At $\xi=0$, this angular velocity becomes $\omega_0=\Big(\frac{2\,M\,a}{b^3+a^2\,b+2\,M\,a^2}\Big)$ and vanishes for $\xi \to \pm\,\infty$ or $r \to \infty$. Thus, for ZAMO, the effective potential of the system becomes
\begin{equation}
    V_{eff}=\frac{\alpha^2}{B}\,\Big[\varepsilon+\frac{K\,E^2}{(H^2+A\,K)}\Big].
    \label{11}
\end{equation}

From the above analysis, we see that the effective potential of the system (either time-like or null geodesics) are influenced by the topological defect produced by a global monopole charge indicating by the parameter $\alpha$, the rotating parameter $a$ of the wormhole, and the wormhole parameter $b$ including mass $M$ of the objects. 

In below part, we study the matter-energy content associated with this topologically charged rotating SK-wormhole, as the space-time described by the line-element (\ref{a4}) is a non-vacuum solution of the field equation. In fact, we show that the energy density of the matter-energy content satisfies the weak energy condition \cite{SWH} even at the wormhole throat, $\xi=0$. This ensures that the considered space-time (\ref{1}) or (\ref{a4}) is an example of a topologically charged rotating wormhole without exotic matter.

\subsection{\bf The matter-energy content and the weak energy condition}

In this part, we discuss the matter contents for this rotating wormhole metric (\ref{a4}) and then the energy conditions. The non-zero components of the Ricci tensor $R_{\mu\nu}$ for the space-time (\ref{a4}) are given by
\begin{eqnarray}
    &&R_{tt}=\frac{4a^2(-1 + \alpha^2)Mr\Big\{9a^2-16Mr+4r^2+4\Big(2a^2+r(-4M + 3r)\Big)\cos 2\theta- a^2\cos 4\theta\Big\}}{2^4\,\Big(r^2 + a^2\cos^2\theta\Big)^4},\nonumber\\
    &&R_{rr}=-\frac{a^2(-1 + \alpha^2)\Big\{7a^2 + 4r^2 + 4(2a^2 + 3r^2)\cos 2\theta+a^2\cos 4\theta\Big\}}{2^3\alpha^2\Big(a^2 + r(-2 M + r)\Big)\Big(r^2 + a^2\cos^2 \theta\Big)^2},\nonumber\\
    &&R_{\theta\theta}=-\frac{2(-1+\alpha^2)\Big\{a^4 + 2r^4 + a^2r(-4M + r) +a^2\Big(a^2 + r(-4M + 3r)\Big)\cos 2\theta\Big\}}{2^2\Big(r^2 + a^2\cos^2 \theta\Big)^2},\nonumber\\
    &&R_{t\phi}=-\frac{16a(-1+\alpha^2)Mr\Big\{3a^4 - 2r^4 + a^2r(-4M + r) +a^2\Big(3a^2 + r(-4M + 3r)\Big)\cos 2\theta \Big\}\sin^2 \theta}{2^4\Big(r^2 + a^2\cos^2 \theta\Big)^4},\nonumber\\
    &&R_{\phi\phi}=-\frac{(-1+\alpha^2)}{2^5\Big(r^2 + a^2\cos^2 \theta\Big)^4}\Big[\Big\{10a^8 - 68a^6Mr + 46a^6r^2 + 32a^4M^2r^2 - 88a^4M r^3+84a^4r^4  \nonumber\\
    &&+ 80a^2r^6 + 32r^8+a^2\Big(15a^6 - 32a^2(5M - 3r)r^3 + 16r^5(-8M + 3r)+a^4r(-62M + 63r)\Big)\cos 2\theta\nonumber\\
    &&+2a^4\Big(3a^4 + a^2r(2M + 9r) + 2r^2(-8M^2 - 2Mr + 3r^2)\Big)\cos 4\theta+a^8\cos 6\theta - 2a^6Mr\cos 6 \theta \nonumber\\
    &&+a^6r^2\cos 6\theta \Big\}\sin^2 \theta\Big].
    \label{b4}
\end{eqnarray}
And the Ricci scalar $R=g_{\mu\nu}\,R^{\mu\nu}$ is given by
\begin{eqnarray}
    R=-\frac{2(-1+\alpha^2)\Big\{7a^4 - 8a^2Mr + 8a^2(r^2 + r^2)+8a^2\Big(a^2 + r(-M + 2r)\Big)\cos 2\theta+ a^4\cos 4\theta\Big\}}{8(r^2 + a^2\cos^2 \theta)^3}.
    \label{b5}
\end{eqnarray}

From the above analysis, we see that the non-zero components of the Ricci tensor $R_{\mu\nu}$ and the Ricci scalar $R$ are all vanishes in the limit $\alpha \to 1$ otherwise non-zeros for $\alpha \neq 1$. Thus, in the limit $\alpha \to 1$, this rotating space-time (\ref{1}) or (\ref{a4}) becomes a vacuum solution of the Einstein's field equations, $R_{\mu\nu}=0$. In this research work, we are mainly interested on analysing the effects of the global monopole charge, where $\alpha \neq 1$ in this rotating wormhole. Below, we show that the global monopole charge influences the curvature properties and hence, changes the physical properties of this rotating wormhole space-time.


The quadratic Ricci invariant $\mathcal{I}=R^{\mu\nu}\,R_{\mu\nu}$ is given by
\begin{eqnarray}
    &&\mathcal{I}=\frac{(-1+\alpha^2)^2}{2^6(r^2+a^2\cos^2\theta)^6}\Big[123a^8 + 128r^8 + 192a^2r^5(-2M + r)+8a^6r(-50M + 39r)\nonumber\\
    &&+384a^4r^2(3M^2- 2Mr + r^2)+4a^2\Big\{45a^6 + 80r^5(-2M + r) + a^4r(-130M + 123r)\nonumber\\
    &&+16a^2r^2(24M^2 - 16Mr + 7r^2)\Big\}\cos 2\theta+4a^4\Big\{17a^4 + 2a^2r(-14M + 25r)\nonumber\\
    &&+16r^2(6M^2-4Mr +3r^2)\Big\}\cos 4\theta+4a^6\Big(3a^2 + r(2M + 5r)\Big)\cos 6\theta+a^8\cos 8\theta \Big].
    \label{bb5}
\end{eqnarray}

The Kretschmann scalar curvature $\mathcal{K}=R^{\mu\nu\lambda\sigma}\,R_{\mu\nu\lambda\sigma}$ is given by
\begin{eqnarray}
    &&\mathcal{K}=\frac{1}{2^6\Big(r^2 + a^2\cos^2 \theta\Big)^6}\Big[83a^8(-1 + \alpha^2)^2+128r^6\Big\{12b^4M^2-4b^2(-1 + \alpha^2)Mr \nonumber\\
    &&+ (-1 + \alpha^2)^2r^2\Big\}+128a^2r^4\Big\{ -10\alpha^2(7 + 2\alpha^2)M^2 - 2(2 - 3\alpha^2 + \alpha^4)Mr + (-1 + \alpha^2)^2r^2\Big\}\nonumber\\
    &&+ 32a^4r^2\Big\{6(11 + 20\alpha^2 + 14\alpha^4)M^2 - 2(13 - 15\alpha^2 + 2\alpha^4)Mr+5(-1 + \alpha^2)^2r^2\Big\}\nonumber\\
    &&+16a^6\Big\{-30\alpha^2M^2 + 2,(1 + 19\alpha^2 - 20\alpha^4)Mr + 9(-1 + \alpha^2)^2r^2\Big\}\nonumber\\
    &&+8a^2\Big\{15a^6(-1 + \alpha^2)^2 +16r^4\Big(-10\alpha^2(7 + 2\alpha^2)M^2 +2(-4 + 3\alpha^2 + \alpha^4)Mr + (-1 + \alpha^2)^2r^2\Big)\nonumber\\
    &&+16a^2r^2\Big((22 + 40\alpha^2 + 28\alpha^4)M^2-2(3 - 5\alpha^2 + 2\alpha^4)Mr + (-1 + \alpha^2)^2r^2\Big)\nonumber\\
    &&+a^4\Big(-90\alpha^2M^2 + 2(1 + 49\alpha^2-50\alpha^4)Mr +31(-1 + \alpha^2)^2r^2\Big)\Big\}\cos 2\theta\nonumber\\
    &&+4a^4\Big\{11a^4 (-1 + \alpha^2)^2+8r^2\Big((22 + 40\alpha^2+28\alpha^4)M^2+2(1 + 5b^2-6\alpha^4)Mr+3(-1 + \alpha^2)^2r^2\Big)\nonumber\\
    &&+4a^2\Big(-18\alpha^2M^2-2(1-5\alpha^2 + 4\alpha^4)Mr+7(-1 + \alpha^2)^2r^2\Big)\Big\}\cos 4\theta\nonumber\\
    &&+8a^6\Big\{a^2(-1 + \alpha^2)^2+r(-2M+r)+\alpha^4r(4M + r)-2\alpha^2(3M^2+Mr+r^2)\Big\}\cos 6\theta\nonumber\\
    &&+a^8(-1 + \alpha^2)^2\cos 8\theta\Big].
    \label{c5}
\end{eqnarray}
One can easily verify that these scalar curvatures obtained above becomes finite at $\xi=0$ and vanishes for $\xi \to \pm\,\infty$. 

Now, we check behaviour of these quantities at the wormhole throat defined by $\xi=0$ in the equatorial plane. At $\xi=0$, that is, $r=b>2\,M$ and in the equatorial plane defined by $\theta=\frac{\pi}{2}$, the components of the Ricci tensor (\ref{b4}) becomes
\begin{eqnarray}
    &&R_{tt}|_{\xi=0,\theta=\pi/2}=-\frac{2\,M\,a^2\,(-1 + \alpha^2)}{b^5},\nonumber\\
    &&R_{rr}|_{\xi=0,\theta=\pi/2}=\frac{a^2\,(-1 + \alpha^2)}{\alpha^2\,b^2\,\Big(a^2 + b\,(-2 M + b)\Big)},\nonumber\\ 
    &&R_{\theta\theta}|_{\xi=0,\theta=\pi/2}=\frac{(-1+\alpha^2)\,(a^2-b^2)}{b^2},\nonumber\\
    &&R_{t\phi}|_{\xi=0,\theta=\pi/2}=\frac{2\,a\,M\,(-1+\alpha^2)(a^2+b^2)}{b^5},\nonumber\\
    &&R_{\phi\phi}|_{\xi=0,\theta=\pi/2}=-\frac{(-1+\alpha^2)\,\Big(b^5+2\,a^4\,M+a^2\,b^2\,(b+4\,M)\Big)}{b^5}.
    \label{c6}
\end{eqnarray}
Also the Ricci scalar, the quadratic Ricci invariant, and the Kretschmann scalar curvature in this case ($\xi=0$ or $r=b>2\,M$, $\theta=\frac{\pi}{2}$) becomes
\begin{eqnarray}
    &&R|_{\xi=0,\theta=\pi/2}=\frac{2\,(a^2-b^2)\,(-1+\alpha^2)}{b^4},\nonumber\\
    &&\mathcal{I}|_{\xi=0,\theta=\pi/2}=\frac{2\,(-1+\alpha^2)^2}{b^8}\,\Big[a^4+b^4-a^2\,b\,(b-2\,M)\Big],\nonumber\\
    &&\mathcal{K}|_{\xi=0,\theta=\pi/2}=\frac{4}{b^8}\,\Big[a^4(-1 +\alpha^2)^2-4a^2bM(-1+\alpha^4)
    +b^2\Big\{12M^2\alpha^4\nonumber\\
    &&-4bM\alpha^2(-1 + \alpha^2)+b^2(-1 + \alpha^2)^2\Big\}\Big].
    \label{c14}
\end{eqnarray}

From the above analysis we see that none of the physical quantities associated with the curvature tensor diverge at the wormhole throat $\xi=0$ considered in the equatorial plane. One can show that the different curvature invariants such as, the Ricci scalar ($R$), the quadratic Ricci invariant ($\mathcal{I}$), and the Kretschamnn scalar ($\mathcal{K}$) all are vanishes at $\xi \to \pm\,\infty$ or $r \to \infty$ and remains finite at $\xi \to 0$. This point is also cleared by analysing the quantity $\frac{1}{(r^2+a^2 \cos^2 \theta)}$ that's remain finite (non-divergence) even at $\theta=\pi/2$ for all values of $r$ since the radial coordinate $r$ lies in the interval $r \in [b>2\,M, \infty)$ instead of $r \in [0, \infty)$.

\begin{figure}
        \centering
        \includegraphics[width=3.2in,height=1.6in]{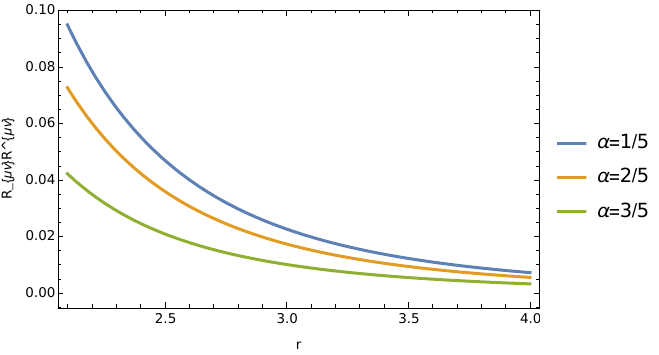}
        \caption{The quadratic Ricci invariant $R^{\mu\nu}\,R_{\mu\nu}$.}
        \label{fig: 1}
        \hfill\\
        \includegraphics[width=3.2in,height=1.6in]{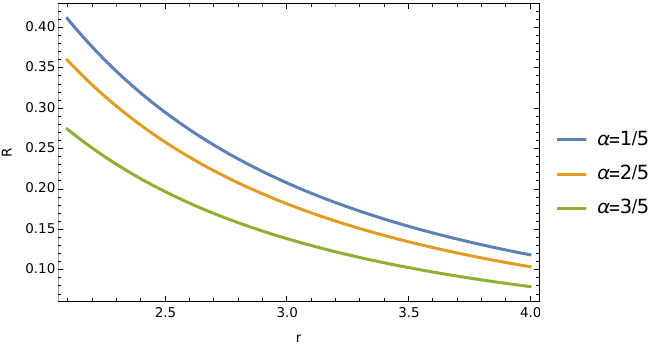}
        \caption{The Ricci scalar $R$.}
        \label{fig: 2}
        \hfill\\
        \includegraphics[width=3.2in,height=1.6in]{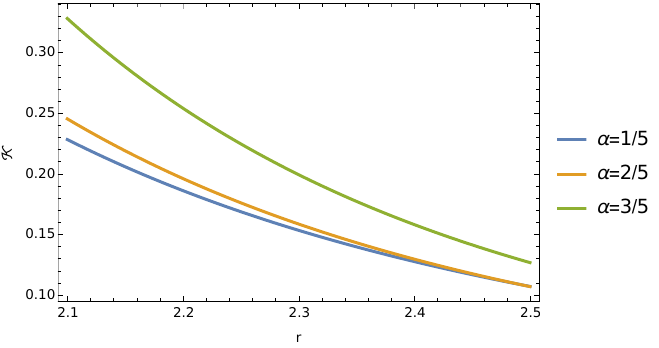}
        \caption{The Kretschmann scalar $\mathcal{K}$.}
        \label{fig: 3}
        \hfill\\
        \includegraphics[width=3.2in,height=1.6in]{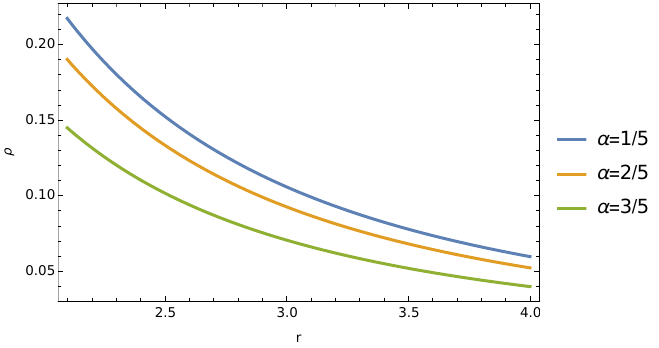}
        \caption{The energy-density}
        \label{fig: 4}
\end{figure}

From the field equations $G^{\mu}_{\,\nu}=R^{\mu}_{\,\nu}-\frac{1}{2}\,\delta^{\mu}_{\,\nu}\,R=T^{\mu}_{\nu}$, we have the energy density
\begin{eqnarray}
    &&\rho=-G^{t}_{t}=-(g^{tt}\,R_{tt}+g^{t\phi}\,R_{\phi t})+\frac{1}{2}\,R\,.
    \label{c8}
\end{eqnarray}
Here $R_{tt}, R_{t\phi}$ are given in (\ref{b4}) and
\begin{eqnarray}
    &&g^{tt}=-\frac{a^4 + 2\,r^4 + a^2\,r\,(2\,M + 3\,r) +a^2\,(a^2 + r\,(-2\,M + r))\,\cos 2\theta}{2\,(r^2+a^2-2\,M \,r)\,(r^2 + a^2\,\cos^2\theta)},\nonumber\\
    &&g^{\phi t}=g^{t\phi}=-\frac{2\,M\,a\,r}{(r^2+a^2-2\,M\,r)\,(r^2 + a^2\,\cos^2\theta)}.
    \label{c9}
\end{eqnarray}

At $\xi=0$, that is, $r=b$ in the equatorial plane defined by $\theta=\frac{\pi}{2}$, these metric components becomes
\begin{eqnarray}
    &&g^{tt}|_{\xi=0}=-\Big(1+\frac{a^2}{b^2}-\frac{2\,M}{b}\Big)^{-1}\,\Bigg[1+\frac{a^2}{b^2}\,\Big(1+\frac{2\,M}{b}\Big) \Bigg],\nonumber\\
    &&g^{t\phi}|_{\xi=0}=-\Big(1+\frac{a^2}{b^2}-\frac{2\,M}{b}\Big)^{-1}\,\Big(\frac{2\,M\,a}{b^3}\Big).
    \label{c10}
\end{eqnarray}
Therefore, the energy density at $\xi=0$, that is, $r=b$ in the equatorial plane $\theta=\frac{\pi}{2}$ is given 
\begin{equation}
    \rho|_{\xi=0,\theta=\pi/2}=\Big(\frac{1-\alpha^2}{b^2}\Big)\,\Bigg[1-\frac{a^2}{b^2}\,\Big(1-\frac{2\,M}{b}\Big)\Bigg]>0
    \label{c11}
\end{equation}
which is positive since $b>2\,M$ and the global monopole parameter $0 < \alpha < 1$ in the gravitation and cosmology. Thus, the energy-density of the matter content satisfies the weak energy condition (WEC) provided the condition $a < b/\sqrt{1-\frac{2\,M}{b}}$ must holds good. For non-rotating wormhole case, where the parameter $a=0$, the energy-density becomes $\rho|_{\xi=0,\theta=\pi/2}=\Big(\frac{1-\alpha^2}{b^2}\Big)$ which is similar to the result obtained in Ref. \cite{FA} for a topologically charged non-rotating Klinkhamer-type wormhole space-time.

We have generated a few plots illustrating the scalar quantities (Figures 1-3) tied to the curvature of space-time, along with the energy density (Figure 4), for varying values of the topological defect of global monopole parameter $\alpha$. In this analysis, we've set the rotation parameter $a=1/2$ and the mass parameter $M=1$ in a system of units, focusing on the equatorial plane defined at $\theta=\pi/2$. It's evident that these scalar quantities exhibit a progressive decline as the radial distance $r$ increases, given a specific $\alpha$ value. As we increase the parameter $\alpha$, we observe a downward shift in the levels of the Ricci scalar, the quadratic Ricci invariant, and the energy-density of the matter content. Conversely, when considering the Kretschmann scalar curvature case, its levels experience an upward shift as the global monopole parameter $\alpha$ increases. We also show in Figure 4 that the energy-density of the matter content is positive for all values of $\xi \in (-\infty, +\infty)$ including at the wormhole throat $\xi=0$ considered in the equatorial plane $\theta=\pi/2$.

\section{Conclusions}

Traversable wormhole solutions are often associated with the need for exotic matter, which violates the weak energy condition. However, this requirement for exotic matter can be potentially avoided in modified theories of gravity that incorporate higher curvature corrections into their gravitational actions. In such theories, the presence of these higher-order terms can lead to the violation of the null-energy condition, effectively mimicking the exotic matter contribution necessary for traversable wormholes. For instance, traversable wormholes without the need for exotic matter have been derived in theories like quadratic gravity and Chern-Simons modified gravity. Researchers have explored various avenues within both general relativity and modified gravity theories in attempts to construct traversable wormholes without the use of exotic matter. It is important to note that while some of these wormhole solutions have been identified, many have been found to be inherently unstable. This highlights the complexity and challenges associated with achieving stable traversable wormholes without exotic matter, even in modified gravity frameworks. In the context of general relativity theory, only a handful works on the rotating wormhole solutions of Einstein's field equations are well-known in the literature. Here, we attempt to present one example of a rotating defect wormhole with topological defect produced by a global monopole charge. 

In this paper, we have introduced a novel topologically charged rotating Schwarzschild-Klinkhamer defect wormhole space-time within the framework of general relativity. Our investigation delves into the trajectories of photon rays across various directions, revealing the influence of key parameters such as the spin parameter $a$, the wormhole parameter $b$, and the global monopole charge $\alpha$. Our analysis extends to the study of geodesic motions, particularly in the equatorial plane defined by $\theta=\pi/2$, as well as a special case pertinent to the perspective of a zero angular momentum observer (ZAMO). Examining the physical characteristics of this rotating wormhole, we find that several quantities associated with space-time curvature, including the Kretschmann scalar, the Ricci scalar, and the quadratic Ricci invariant, remain finite even as the wormhole throat ($\xi=0$) vanishes for $\xi \to \pm\,\infty$ or $r \to \infty$. This observation underscores the singularity-free wormhole under consideration. Furthermore, we have undertaken an exploration of the matter-energy content associated in this topologically charged rotating wormhole, as it is a non-vacuum solution of Einstein's field equations. Our findings indicate that the energy density of the matter-energy content adheres to the weak energy condition (WEC). Consequently, this rotating wormhole does not necessitate exotic matter for its stability. 

As a prospective avenue for future research, we propose a comprehensive examination of the deflection angle (gravitational lensing) of photon rays within the context of this stationary and rotating wormhole. This investigation aims to enhance our understanding of the optical properties associated with the wormhole's gravitational field. Additionally, we plan to conduct an analysis of the geometrical properties inherent in this rotating wormhole model. This detailed study will contribute valuable insights into the intricate structure and characteristics of the space-time geometry, further advancing our comprehension of the physical implications and behaviors of the proposed defect wormhole.

\global\long\def\link#1#2{\href{http://eudml.org/#1}{#2}}
 \global\long\def\doi#1#2{\href{http://dx.doi.org/#1}{#2}}
 \global\long\def\arXiv#1#2{\href{http://arxiv.org/abs/#1}{arXiv:#1 [#2]}}
 \global\long\def\arXivOld#1{\href{http://arxiv.org/abs/#1}{arXiv:#1}}


\begin{thebibliography}{99}

\bibitem{MV} 
M. Visser, {\it Lorentzian Wormholes: From Einstein to Hawking}, American Institute of Physics, Melville, NY (1996).

\bibitem{FRK} 
F. R. Klinkhamer, \href{https://doi.org/10.5506/APhysPolB.54.5-A3}{Acta Phys. Polon. B {\bf 54}, 5-A3 (2023)} [arXiv:2301.00724 [gr-qc]].

\bibitem{FRK3} 
F. R. Klinkhamer, \href{https://doi.org/10.5506/APhysPolB.54.7-A3}{Acta Phys. Polon. B {\bf 54}, 7-A3 (2023)} [arXiv:2305.13278 [gr-qc]].

\bibitem{FA} 
F. Ahmed, \href{https://doi.org/10.1088/1475-7516/2023/11/010}{JCAP 11 ({\bf 2023}) 010} [arXiv: 2307.08503 [gr-qc]].

\bibitem{FA2} 
F. Ahmed, \href{https://doi.org/10.1088/1475-7516/2023/11/082}{JCAP 11 ({\bf 2023}) 082} [arXiv: 2308.00012 [gr-qc]].

\bibitem{JCF} 
J. C. Feng, \href{https://doi.org/10.1088/1361-6382/acf2de}{Class. Quantum Grav. {\bf 40}, 197002 (2023)} [arXiv: 2308.11885 [gr-qc]].

\bibitem{BGV}
J. Baines, R. Gaur and M. Visser, \href{https://doi.org/10.3390/universe9100452}{Universe {\bf 2023}, 9(10), 452} [arXiv: 2308.16624 [gr-qc]].

\bibitem{ET} 
E. Teo, \href{https://doi.org/10.1103/PhysRevD.58.024014}{Phys. Rev. {\bf D 58}, 024014 (1998)} [arXiv:gr-qc/9803098].

\bibitem{VMK} 
V. M. Khatsymovsky, \href{https://doi.org/10.1016/S0370-2693(98)00448-1}{Phys. Lett. {\bf B 429}, 254 (1998)} [arXiv:gr-qc/9803027].

\bibitem{ss1} 
G. Clément and D. Gal'tsov, \href{https://doi.org/10.1016/j.physletb.2023.137677}{Phys. Lett. B {\bf 838}, 137677 (2023)} [arXiv:2210.08913 [gr-qc]].

\bibitem{ss2} 
X. Y. Chew, B. Kleihaus, J. Kunz, V. Dzhunushaliev and V. Folomeev, \href{https://doi.org/10.1103/PhysRevD.100.044019}{Phys. Rev. {\bf D 100}, 044019 (2019)} [arXiv:1906.08742 [gr-qc]].

\bibitem{ss3} 
E. Caceres, A. S. Misobuchi and M.-L Xiao, \href{https://doi.org/10.1007/JHEP12(2018)005}{JHEP {\bf 2018}, 5 (2018)} [arXiv:1807.07239 [hep-th]].

\bibitem{ss4} 
M. S. Volkov, \href{https://doi.org/10.1103/PhysRevD.104.124064}{Phys. Rev. {\bf D 104}, 124064 (2021)} [arXiv:2109.14496 [gr-qc]].

\bibitem{ss5} 
B. Kleihaus and J. Kunz, \href{http://dx.doi.org/10.1103/PhysRevD.90.121503}{Phys. Rev. {\bf D 90}, 121503(R) (2014)} [arXiv:1409.1503 [gr-qc]].

\bibitem{hh1} 
M. Azreg-Aïnou, \href{https://doi.org/10.1140/epjc/s10052-014-2865-8}{Eur. Phys. J. C {\bf 74}, 2865 (2014)} [arXiv:1401.4292 [gr-qc]].

\bibitem{hh2} 
M. Azreg-Aïnou, \href{https://doi.org/10.1103/PhysRevD.90.064041}{Phys. Rev. {\bf D 90}, 064041 (2014)} [arXiv: 1405.2569 [gr-qc]].

\bibitem{hh3} 
M. Azreg.-Aïnou, \href{https://doi.org/10.1016/j.physletb.2014.01.041}{Phys. Lett. B {\bf 730}, 95 (2014)} [arXiv: 1401.0787 [gr-qc]].

\bibitem{hh4} 
M. Azreg-Aïnou, \href{https://doi.org/10.1140/epjc/s10052-015-3836-4}{Eur. Phys. J. C {\bf 76}, 7 (2016)} [arXiv: 1509.00234 [gr-qc]].

\bibitem{hh5} 
A. Cisterna, K. Müller, K. Pallikaris and A. Viganò, \href{https://doi.org/10.1103/PhysRevD.108.024066}{Phys. Rev. {\bf D 108}, 024066 (2023)} [arXiv: 2306.14541 [gr-qc]].

\bibitem{kk1} 
T. W. B. Kibble, \href{https://doi.org/10.1088/0305-4470/9/8/029}{J. Phys. A: Math. \& Gen. {\bf 9}, 1387 (1976).}

\bibitem{kk2} 
T. W. B. Kibble, \href{https://doi.org/10.1016/0370-1573(80)90091-5}{Phys. Rept. {\bf 67}, 183 (1980).} 

\bibitem{kk3}  
T. W. B. Kibble, \href{https://www.actaphys.uj.edu.pl/R/13/10/723}{Acta Phys. Polon. {\bf B 13}, 723 (1982).}

\bibitem{AV} 
A. Vilenkin and E. P. S. Shellard, {\it Strings and Other Topological Defects}, Cambridge University Press, Cambridge (1994).

\bibitem{kk51} 
R. Brandenberger, \href{https://doi.org/10.1142/S0217751X8700003X}{Int. J. Mod. Phys. {\bf A 02}, 77 (1987).}

\bibitem{kk52} 
R. Brandenberger, L. Perivolaropoulos and A. Stebbins, \href{https://doi.org/10.1142/S0217751X9000074X}{Int. J. Mod. Phys. {\bf A 05}, 1633 (1990)}.

\bibitem{kk5} 
R. Brandenberger, \href{https://doi.org/10.1142/S0217751X9400090X}{Int. J. Mod. Phys. {\bf A 9}, 2117 (1994)} [arXiv:astro-ph/9310041]. 

\bibitem{kk6} 
M. Hindmarsh and T. W. B. Kibble, \href{https://doi.org/10.1088/0034-4885/58/5/001}{Rept. Prog. Phys. {\bf 58}, 477 (1995)} [arXiv:hep-ph/9411342].

\bibitem{kk7}  
S. Sarangi and S. H. H. Tye, \href{https://doi.org/10.1016/S0370-2693(02)01824-5}{Phys. Lett. {\bf B 536}, 185 (2002)} [arXiv:hep-th/0204074]. 

\bibitem{kk8} 
R. Jeannerot, J. Rocher and M. Sakellariadou, \href{https://doi.org/10.1103/PhysRevD.68.103514}{Phys. Rev. {\bf D 68}, 103514 (2003)} [arXiv:hep-ph/0308134]. 

\bibitem{kk9} 
G. Dvali and A. Vilenkin, \href{https://doi.org/10.1088/1475-7516/2004/03/010}{JCAP 03 ({\bf 2004}) 010} [arXiv:hep-th/0312007].

\bibitem{kk12} 
D. Harari and C. Loust’o, \href{https://doi.org/10.1103/PhysRevD.42.2626}{Phys. Rev. {\bf D 42}, 2626 (1990).}

\bibitem{kk13} 
A. Vilenkin, \href{https://doi.org/10.1103/PhysRevLett.72.3137}{Phys. Rev. Lett. {\bf 72}, 3137 (1994)} [arXiv:hep-th/9402085].

\bibitem{kk14}
R. Basu, A. H. Guth and A. Vilenkin, \href{https://doi.org/10.1103/PhysRevD.44.340}{Phys. Rev. {\bf D 44}, 340 (1991).}

\bibitem{kk15}
R. Basu and A. Vilenkin, \href{https://doi.org/10.1103/PhysRevD.50.7150}{Phys. Rev. {\bf D 50}, 7150 (1994)} [arXiv:gr-qc/9402040].

\bibitem{kk16} 
P. Das and M. Kalam, \href{https://doi.org/10.1140/epjc/s10052-022-10322-z}{Eur. Phys. J C {\bf 82}, 342 (2022).}

\bibitem{RPK} 
R. H. Boyer and R. W. Lindquist, \href{https://doi.org/10.1063/1.1705193}{J. Math. Phys. {\bf 8}, 265 (1967).}

\bibitem{SWH} 
S. W. Hawking and G. F. R. Ellis, {\it The Large Scale Structure Of Space–Time}, Cambridge University Press, Cambridge (1975).


\end{thebibliography}
\end{document}